\documentstyle[graphicx,prb,aps]{revtex}
\def\beq{\begin{eqnarray}}
\def\eeq{\end{eqnarray}}
\begin{document}
\author{M.R. Cimberle, C. Ferdeghini, E. Giannini, D. Marr\'e, M. Putti, 
A. Siri}
\address{INFM / CNR, Dipartimento di Fisica, Universit\`a di Genova,\\
via Dodecaneso 33, Genova 16146, Italy}
\author{F. Federici, A. Varlamov}
\address{Laboratorio ``Forum" dell'INFM, Dipartimento di Fisica \\
Universit\`a di Firenze, Largo E.Fermi 2,  50125 Firenze, Italy} 
\title{The crossover between Aslamazov-Larkin and short wavelength
fluctuation regimes \\
in HTS conductivity experiments}
\date{\today}
\maketitle

\begin{abstract}

We present paraconductivity (AL) measurements in three different high
temperature superconductors: a melt textured $YBa_2Cu_3O_7$ sample,
a $Bi_2Sr_2CaCu_2O_8$ epitaxial thin film and a highly textured
$Bi_2Sr_2Ca_2Cu_3O_{10}$ tape. 
The crossovers between different temperature regimes in excess 
conductivity have been analysed. 
The Lawrence-Doniach (LD) crossover, which separates the 2D and 3D
regimes, shifts from lower to higher temperatures as the compound
anisotropy decreases.
Once the LD crossover is overcome, the fluctuation conductivity of the
three compounds shows the same universal
behaviour: for $\epsilon =\ln T/T_{\rm c}> 0.23$ all the
curves bend down according to the $1/\epsilon^3$ law. 
This asymptotic behaviour 
was theoretically predicted previously for the high temperature
region where the  short wavelength fluctuations (SWF) become important. 

\end{abstract}


It is well known that, owing to strong anisotropy,  
high critical temperature and low charge carrier concentration, 
thermodynamic fluctuations play an important role in the explanation of  
the normal state properties of high temperature superconductors (HTS).
Just after the realization of high quality epitaxial single crystal 
samples, the in-plane fluctuation conductivity was investigated in detail 
and the Lawrence-Doniach (LD) crossover between three-dimensional (3D) 
and two-dimensional (2D) regimes (or at least a tendency to it) was
observed in the vicinity of $T_{\rm c}$ in the majority of HTS 
compounds. Analogous phenomena were observed in magnetic susceptibility,
thermoconductivity \cite{Aus} and other properties of HTS.

Let us recall that LD crossover takes place in the temperature dependence of
in-plane conductivity and it is related to the fact that fluctuative
Cooper pairs motions change from 2D to 3D rotations.
It takes place at the temperature $T_{\rm LD}$ which is defined by the 
condition $\xi_{\rm c}(T_{\rm LD})\approx s$, where $\xi_{\rm c}$ is the
coherence length and $s$ is the interlayer distance. Nevertheless, 
the LD crossover in the temperature dependencies of different characteristics does not
exhaust all possibilities: additional crossovers can be observed in HTS compounds. 
For instance, another kind of crossover (0D $\rightarrow$ 3D) can take place in 
$c$-axis paraconductivity temperature dependence, at the same temperature $T_{\rm LD}$. 
It is due to the fact that the pair propagation along c-axis
has a zero-dimensional character relatively far from $T_{\rm c}$ and it changes into
a three-dimensional rotation in the immediate vicinity of the transition.
This effect was predicted \cite{Klemm} and observed \cite{Buz} in
fully oxygenated YBCO samples while in BSCCO samples it is masked
by the increase of resistivity due to fluctution density of states
renormalization. Below we will remind the reader of the possible kinds of
crossover phenomena taking place in layered superconductors and finally we
will present the experimental evidences of the crossover related to the
breakdown of Ginzburg Landau (GL) approximation, due to the importance of
short wavelength fluctuations. 

How the LD crossover appears in the framework of the GL theory can be shown explicitly
considering the model of an open electron Fermi surface which, 
for instance, can be chosen in the form of a ``corrugated cylinder" \cite{note1}. 
In this case the energy spectrum has the form
\beq
\xi({\bf p}) =\epsilon_0({\bf p})+J\cos(p_{\perp}s)-E_{\rm F},\label{corrug}
\eeq
where $\epsilon_0({\bf p})={\bf p}^2/(2m)$, $p\equiv({\bf p}, p_{\perp})$, 
${\bf p}\equiv(p_x,p_y)$ is a two-dimensional, intralayer wavevector, 
and $J$ is an effective hopping energy. 
The Fermi surface is defined by the condition $\xi(p_{\rm F})=0$ and 
$E_{\rm F}$ is the Fermi energy. This spectrum is 
the most appropriate for strongly anisotropic layered materials
where $J/E_{\rm F}\ll 1$. 

In the framework of the Ginzburg-Landau theory for an isotropic spectrum, 
the fluctuation contribution to the free energy of a superconductor
above the critical temperature can be presented as the sum 
over long wavelength fluctuations \cite{abr}:
\beq
F = - T \sum_{\bf q} \ln{{ \pi T} \over {\alpha(\epsilon +
 \eta {\bf q}^2)}}.
\label{F1}
\eeq
where $\alpha$ and $\eta$ are the coefficients of GL theory. 
In the expression (\ref{F1}) 
$\epsilon=\ln(T/T_{\rm c})\simeq (T-T_{\rm c})/T_{\rm c}
\ll 1$ is supposed.
The term $\eta {\bf q}^2$ in (\ref{F1})  
is actually the result of angular averaging over the isotropic Fermi surface
of $({\bf v}\cdot{\bf q})^2$ (up to some dimensional coefficient)
and in the case of anisotropic
spectrum (\ref{corrug}) must be substituted by the more sofisticated
expression including the additional dependence on $q_{\perp}$:
\beq
\left\langle\left({\bf v }\cdot{\bf q}\right)^2\right\rangle=
\left\langle\left[\xi\left({\bf p}\right)-\xi\left({\bf q}-
{\bf p}\right)\right]^2\right\rangle=\frac12\left(v_{\rm F}^2{\bf q}^2+
4J^2\sin^2(q_{\perp} s/2)\right)\label{D}
\eeq

Formally the crossover temperature is defined by the analysis of the 
denominator in the expression (\ref{F1}).
Evidently, in the strictly 2D case, the hopping integral 
$J\rightarrow 0$ and 
the integration over $q_{\perp}$ is reduced to the redefinition of the 
density of states only.
The crossover in the fluctuation behaviour takes place when the
integration over the transverse variable becomes important and changes 
the temperature dependence of the appropriate fluctuation correction. 
This happens, evidently, when the transversal part reaches the 
reduced temperature $\epsilon$ or, in other words, when
$\xi_{\rm c}(T_{\rm LD})\approx s$. 
For not too anisotropic compounds, such crossover
can be delayed up to temperatures high enough, far from the immediate 
vicinity of the transition. 

Analogous phenomena can be also observed in the presence of
a magnetic and even electric field \cite{VR}. The reason for this fact can 
be understood from the following consideration. The integrand
function in (\ref{F1}), in its exact formulation, is nothing else than
the fundamental solution of the Maki-De Gennes equation
(see, for instance, the review article of K. Maki in \cite{Parks}).  
When some external field is applied, the eigen-value of the 
unperturbed Hamiltonian is renormalized and, starting from some value 
of the interaction strength, the pole in $q$-dependence is determined 
more by the renormalization of the eigen-function than by the reduced 
temperature $\epsilon$.

Further, the presence of an {\bf ac} electromagnetic field leads to
the appearance of $-i\omega$ side by side with the energy eigen-value
in the Maki-De Gennes equation \cite{Parks}. This results in a possible
crossover in the frequency dependence of fluctuation conductivity
\cite{Sch,AV,FV}.

Another kind of crossover is related to the geometry of the samples.
Let us suppose that the epitaxial film of layered 
superconductor consists of several layers only. In this case, very
near to $T_{\rm c}$, the transverse size of Cooper pairs
fluctuations can exceed the film thickness and the class of possible
pairs motions is again reduced to that of 2D rotations \cite{VY}.
This kind of crossover cannot be easily observed in HTS films
because of the extremely short coherence length, but it becomes
observable for conventional superconducting thin films or
in artificial superlattices of materials with large coherence length.

The last kind of crossover, which we would like to discuss here, is
related to the breakdown of the GL approach
in describing the fluctuations relatively far from the transition
where the assumption of the domination of long wavelength fluctuations
contribution is no longer valid. The generalization of the fluctuation 
conductivity theory for high temperature region has been already analysed
\cite{rvv}.
Namely the short wavelength fluctuations have been taken into account
and the universal formula for paraconductivity has been obtained
\beq
\label{al}
\sigma_{\rm fl}^{\rm 2D}=\frac{e^2}{16\hbar s}f(\epsilon).
\eeq
In the GL region of temperature, where $\epsilon\ll 1$, 
$f(\epsilon)=1/\epsilon$ and the result coincides with the well known
AL one. In the opposite case $\epsilon\gg 1$, for clean
2D superconductors, $f(\epsilon)\sim 1/\epsilon^3=1/\ln^3(T/T_{\rm c})$
was carried out. In the theoretical consideration it was natural to 
assume formally the very rigid
restriction $\epsilon\gg 1$ for the validity of the latter asymptotic behaviour.
Nevertheless, as it will be seen below, in experiments 
the crossover to this asymptotic behaviour takes place universally for all
the samples investigated at $\epsilon \sim 0.23$ and this can be
attributed to some particularly fast convergence of the integrals in the
expression of $f(\epsilon)$.

The long tails in the in-plane fluctuation conductivity of HTS materials
have been observed frequently. One of the efforts to fit the high
temperature paraconductivity with the extended AL theory results
was undertaken in \cite{bmmrvv} where  the deviation of the
excess conductivity from AL behaviour was analysed for three
$Bi_2Sr_2CaCu_2O_8$ epitaxial films.
Very good fit with the formula (\ref{al}) was found in the region of
temperatures $0.02\lesssim\epsilon\lesssim 0.14$. 
We show here that the careful analysis of the higher temperature
region (just above the edge of the region investigated in \cite{bmmrvv})
allows to observe the surprisingly early approaching to the
SWF asymptotic regime (at the reduced teperature 
$\epsilon^{\ast}\sim\ln (T^{\ast}/T_{\rm c})\sim 0.23$).

We have performed resistivity
measurement of three different HTS compounds: a melt textured 
YBa$_{2}$Cu$_{3}$O$_{7}$ sample (Y123), a Bi$_{2}$Sr$_{2}$CaCu$_{2}$O$_{8}$  
(Bi2212) thick film and a highly textured Bi$_{2}$Sr$_{2}$Ca$_{2}$Cu$_{3}$O$_{10}$ 
(Bi2223) tape. 
The Y123 was obtained by melting \cite{1}; the sample was cut in a
nearly regular parallelepipedal shape with a cross section of about
$4~mm^{2}$ and a length of $7~mm$. The resistivity measurements were
performed from 85 to  $330~K$. The critical temperature, defined as the
point where the temperature derivative is maximal, is $92~K$;
$\rho_{\rm N}(100 K)=120~\mu\Omega cm$, where $\rho_{\rm N}$ is the 
resistivity in the normal state extrapolated from the high temperature
region where $\rho$ is linear. The Bi2212 film was 
prepared by a liquid phase epitaxy technique \cite{2}. 
The film has a thickness of about  $1~\mu m$. The resistivity
measurements were performed from 80 to $170~K$. The critical temperature was
estimated to be $84.2~K$ and $\rho_{\rm N}(100~K)=150\mu\Omega cm$.
The Bi2223 tape was obtained by means of the power in tube procedure, 
as described elsewhere \cite{3}.
The thickness of the oxide filament inside the tape was about $30~\mu m$;
the filament turned out to be strongly textured (rocking angle
$\approx 8^{\circ} $) with the $c$-axis oriented perpendicular to the tape 
plane. The resistivity measurements were performed in the range from 100 to
$250~K$, after removing the silver sheathing chemically.
The critical temperature was estimated to be $108~K$ and $\rho_{\rm N} 
(100 K) = 300~\mu\Omega cm$. 
We ascribe this high value of $\rho_{\rm N}$ to different causes: first, 
the grain boundaries may determine a resistance in series with the grain
resistance; second, the chemical treatment may have damaged the surface
of the sample and the effective cross section of the superconductor
can be decreased. 

The excess conductivity was estimated by subtracting the background of the
normal state conductivity $\sigma_{\rm N} =1/\rho_{\rm N}$. The
evaluation of $\rho_{\rm N}$ was made with particular accuracy; in fact, 
starting the interpolation at a certain temperature corresponds to forcing 
$\sigma_{fl}$ to vanish artificially at such temperature. 
Therefore, we need to estimate $\rho_{\rm N}$ at a temperature as large
as possible and to verify that $\rho_{\rm N}$ does not depend on 
the temperature range where the interpolation is performed.
In the case of Y123 sample the resistivity shows a linear behavior
from 160 to $330~K$. In this range we have verified that $\rho_{\rm N}$ 
does not change by shifting the interpolation temperature region. 

Therefore, for the Y123 sample, the upper limit of $\epsilon$ at which
the excess conductivity may be analysed is 
$\epsilon_{\rm up}\approx ln(160/92)=0.55$. In an analogous way we obtain 
$\epsilon_{\rm up}\approx 0.46$ and 0.51 for Bi2212 and Bi2223, respectively.

In Fig. 1, in a log-log scale, we plot 
$\sigma_{\rm fl}\left(\displaystyle{\frac{16hs}{2\pi e^2}}\right)$ 
as a function of $\epsilon$ for the three samples; the solid line 
represents $1/\epsilon$, the dashed line $0.055/\epsilon^{3}$ and the dotted
line is $3.2/\sqrt{\epsilon}$. The interlayer distance $s$ is considered as
a free parameter and it has been adjusted so that the experimental data can
follow the $1/\epsilon$ behaviour in the $\epsilon$ region where the AL
behavior is expected.

We can see that all the curves exhibit the same general behaviour. 
The region where the 2D $1/\epsilon$ behaviour is followed, has different
extension for each compound, depending on its anisotropy, and at
$\epsilon\approx 0.23$ all the curves bend downward and follow the same
asymptotic $1/\epsilon^3$ behaviour.

We discuss now some features in detail:

\begin{itemize}
\item[1)]  The interlayer distance values we find are the following: 
for Y123 we obtain $s\approx 13~\AA$ which must be compared with the 
YBCO interlayer distance that is about $12~\AA$; for Bi2212
we obtain $s\approx 11~\AA$ to be compared with $15~\AA$, and for Bi2223
we obtain $s\approx 25~\AA$ to be compared with $18~\AA$. 
The differences in the interlayer distance evaluation are all compatible
with the uncertainty on the geometrical factors. We point out that the 
smallest error is for Y123 (about $10\%$) that is a bulk sample with a
well defined geometry. Larger errors are found for the Bi2212 thick film 
(about $30\%$), for which the evaluation of the thickness is rough, 
and for the Bi2223 tape (about $40\%$) for which an overestimation
of the cross section of the tape is possible, as we mentioned above.
We conclude that the AL behaviour is well followed. 

\item[2)] On the low $\epsilon$ value side $(\epsilon <0.2)$ the three
compounds show different behaviours due to the different extension of
the AL region. The least anisotropic compound, Y123, for $\epsilon <0.1$
bends going asymptotically to the 3D behaviour ($1/\epsilon^{0.5}$) showing
the LD crossover at $\epsilon\approx 0.09$; the Bi2223 sample
starts to bend for $\epsilon <0.03$ while the most anisotropic Bi2212 in the 
overall $\epsilon$ range considered shows the 2D behaviour.

\item[3)]  On the high $\epsilon$ value side, starting from the AL behaviour, 
the curves show a crossover at about $\epsilon = 0.23$ and then bend downward
following the asymptotic $1/\epsilon^{3}$ behaviour.
At the value $\epsilon\approx 0.45$ all the curves drop indicating the end of
the observable fluctuation regime. This value is lower than the above
reported $\epsilon_{\rm up}$ values, at which the fluctuation conductivity
comes out to be zero.
\end{itemize}

To conclude: we have observed in three different HTS compounds the 
universal high temperature behaviour of the in-plane conductivity
that manifests itself in the 2D regime, once the LD crossover is passed.
Beyond the AL regime all the curves reach soon the SWF
$1/\epsilon^{3}$  regime. For all the compounds the crossover occurs at the
same point $\epsilon\approx 0.23$, which corresponds to
$T\approx 1.3 T_{\rm c}$ and, therefore, is experimentally well observable.
The universality of the paraconductivity behaviour is much more surprising if
we consider that it has been observed in three compounds with different
crystallografic structure and anisotropy, and moreover prepared by means of
very different techniques.

We gratefully acknowledge the fruitful discussions with Giuseppe Balestrino.

\newpage
\begin{figure}
\caption{$\sigma_{\rm fl}\left(\displaystyle{\frac{16hs}{2\pi e^2}}\right)$
 vs $\epsilon$  for Y123 (triangle), Bi2212 (square) and  Bi223 (circlet);
the solid line is $1/\epsilon$, the dashed line $0.055/\epsilon^{3}$, and the dotted
line is $3.2/\sqrt{\epsilon}$}
\end{figure}


\begin{thebibliography}{99}
\bibitem{note1} In principle it is possible to
start from ellipsoidal Fermi surface, but the crossover in this 
case can be observed when the increase of anisotropy results
 in the intersection of the Fermi surface with the Brillouin 
zone boundary only. This fact implies actually the return to the open Fermi
surface.
\end{thebibliography}
\end{document}